\documentstyle[12pt]{article}
\def\a{\alpha}
\def\b{\beta}
\def\c{\chi}
\def\d{\delta}
\def\e{\epsilon}

\def\g{\gamma}

\def\p{\psi}
\def\k{\kappa}
\def\l{\lambda}
\def\m{\mu}
\def\n{\nu}
\def\o{\omega}

\def\s{\sigma}

\def\G{\Gamma}

\def\L{\Lambda}

\def\vf{\varphi}

\def\bp{\bar{\p}}

\def\be{\begin{equation}}
\def\ee{\end{equation}}
\def\arr{\begin{array}{rll}}
\def\ea{\end{array}}
\def\bea{\begin{eqnarray}}
\def\eea{\end{eqnarray}}

\begin{document}

\begin{titlepage}
\noindent

\vskip 2cm

\begin{center}

{\Large{\bf The geometry of $N=4$ twisted string} }\\

\vskip 2cm

{\large Stefano Bellucci}~\footnote{
E-mail: Stefano.Bellucci@lnf.infn.it}${}^a$ ,
{\large Alexei Deriglazov}~\footnote{
On leave from Department of Mathematical Physics,
Tomsk Polytechnical University, Tomsk, Russia\\
\phantom{XX}
E-mail: alexei@fisica.ufjf.br}${}^b$ and
{\large Anton Galajinsky}~\footnote{
On leave from Department of Mathematical Physics,
Tomsk Polytechnical University, Tomsk, Russia\\
\phantom{XX}
E-mail: agalajin@lnf.infn.it,
galajin@mph.phtd.tpu.edu.ru}${}^a$

\vskip 0.4cm

${}^a$ {\it   INFN--Laboratori Nazionali di Frascati, C.P. 13,
00044 Frascati, Italy}

\vskip 0.4cm

${}^b${\it  Instituto de Fisica, Universidade Federal de
Juiz de Fora, MG, Brazil}
\vskip 0.4cm

\end{center}

\vskip 1.5cm

\begin{abstract}
\noindent
We compare $N=2$ string and $N=4$ topological string within
the framework of the sigma model approach. Being classically equivalent
on a flat background, the theories are shown to lead to
different geometries when put in a curved space.
In contrast to the well studied
K\"ahler geometry characterising the former case,
in the latter case a manifold has to admit a covariantly constant
holomorphic two--form in order to support an $N=4$ twisted supersymmetry.
This restricts the holonomy group to be a subgroup of $SU(1,1)$
and leads to a Ricci--flat manifold. We speculate that,
the $N=4$ topological formalism is an appropriate framework
to smooth down ultraviolet divergences intrinsic to the
$N=2$ theory.
\end{abstract}

\vspace{0.5cm}


\end{titlepage}

\noindent
{\bf 1. Introduction}\\[-4pt]

\vskip 0.4cm

String theory and nonlinear sigma models are intimately linked to each
other. Having originated from different sources, namely,
the study of dual models of hadrons on the one hand and
the search for renormalizable field theories in $d\ge 2$ on the other,
they have been soon recognised to be connected as the latter provided
deep insights into the former (see e.g.~\cite{tseyt,callan,sen}).
To mention only the most significant points, the gauging of global
(super) symmetries of a nonlinear sigma model typically results in a string
theory (in the Neveu--Schwarz--Ramond formalism)
coupled to background massless modes, while
the one--loop finiteness fixes the effective low energy dynamics
of the string partner. It should be remembered that, since the gauging
brings extra constraints into the formalism (normally forming
an $N$--extended superconformal algebra) the resulting string theory is
not necessarily critical and that point is to be examined on its own.

Parallel to the progress of string theory on the supersymmetry route,
the nonlinear sigma models revealed a number of striking properties
in the supersymmetric area. With the number of global supersymmetries growing,
a background geometry becomes severely restricted. Admitting an automatic
$N=1$ supersymmetric generalisation~\cite{freed1}, the model was shown to
require a K\"ahler geometry in order to support an $N=2$ global
supersymmetry~\cite{gaume1}. The $N=4$ case, which also corresponds
to a maximally extended supersymmetry~\footnote{$N=3$ automatically
implies $N=4$, as a product of two complex structures yields a third
one~\cite{gaume1}.}, appeals to a hyper K\"ahler space, the latter
being automatically Ricci--flat~\cite{gaume1}. Interestingly enough,
the $N\le 4$ bound correlates well with that known for an
$N$--extended superconformal algebra (SCA) admitting a central
extension~\cite{ramsch},
the latter typically underlying a string theory with an $N$-extended
local supersymmetry on the world--sheet.

Gauged $N=2$ nonlinear sigma model, or $N=2$ string (coupled to
background), has attracted considerable interest over the past decade
(for a comprehensive list of references see, for example,
Ref.~\cite{lecht}).
The theory is critical in
two spatial and two temporal dimensions (or a four--dimensional Euclidean
space) and contains the only physical state in the quantum spectrum.
Being a massless scalar, the latter can be associated with
either the K\"ahler potential (closed string)
or the Yang scalar (open string)~\cite{ov}.
Notice, however, that although the $N=2$ model does provide a
satisfactory stringy description
of self--dual gauge theory or self--dual gravity,
a manifest Lorentz invariance is missing and, in spite of being the theory
of an $N=2$, $d=2$ supergravity coupled to matter, the model fails to
produce fermions in the quantum spectrum
(see, however, a recent work~\cite{bgl}).

At the classical level the former drawback has been overcome
recently~\cite{bg1,bg2} based on an earlier $N=4$ topological
formalism by Berkovits and Vafa~\cite{berkvafa}. According to the
$N=4$ topological prescription, one adds to the theory
two more fermionic currents (of conformal spin 3/2) and two more
bosonic ones (of conformal spin 1) which, on the one hand extend
the $N=2$ SCA to a small $N=4$ SCA, but on the other hand do not change
the physical content of the model as they prove to be functionally
dependent. The key point, however, is that this extension brings
an extra ${U(1,1)}_{outer}$ symmetry to the formalism
(see~\cite{bg1,bg2} for an explicit realization),
which thus raises the global symmetry group to that including
the full Lorentz group (recall $SO(2,2)\simeq SU(1,1)\times SU(1,1)'$).
Quantum equivalence of the two approaches has been established
by explicit evaluation of scattering amplitudes\footnote{The $N=4$
topological prescription appeals to a specific topological twist
which does not treat all the currents on equal footing and
breaks the Lorentz group to $U(1,1)$.} in Ref.~\cite{berkvafa}.
Based on the symmetry argument, the $N=4$ topological string action
has been constructed in~\cite{bg1} just by installing the ${U(1,1)}_{outer}$
into the action of the $N=2$ string.
Curiously enough, to a great extent
the situation resembles what happens for
the Green--Schwarz superstring, where extracting an independent
set of fermionic first class constraints is known to be in a conflict with
manifest Lorentz covariance\footnote{For a detailed discussion on 
covariant quantization of the Green-Schwarz superstring and kappa symmetry,
see e.g. \cite{b1}.}.

Given the $N=4$ topological string action, the natural question to ask is:
Which is the geometry of a non--linear sigma model associated to it?
In the present paper we address this issue and show that (i)
The global supersymmetry in the question is actually $N=4$
twisted supersymmetry (that is why
the title of the paper) and (ii) Apart from being a K\"ahler space,
a target manifold has to admit a covariantly constant
holomorphic two form, in order to support the latter.

As is well known (see e.g.~\cite{lichn}), the last point restricts the
holonomy group to be a subgroup of $SU(1,1)$ which is equivalent to
the Ricci--flatness condition for the Riemann 
tensor (some constructive examples of such manifolds can be found
e.g. in Ref.~\cite{ov}). Alternatively,
working in real coordinates, in addition to a covariantly constant
complex structure characterising the K\"ahler geometry, one reveals
two covariantly constant "real" structures (almost product structures),
forming altogether a pseudo quaternionic algebra.
The geometry of such a type is known as a pseudo hyper K\"ahler
geometry~\cite{hit,gibb}
and has been recently discussed in the sigma model context by
Hull~\cite{hull1}, and Abou Zeid and Hull~\cite{hull2}.

Finally, as the one--loop calculation
proceeds along the same lines both for the $N=2$ nonlinear sigma
model and for the $N=4$ twisted generalisation
(requiring a Ricci--flat manifold for the one--loop
ultraviolet finiteness~\cite{mukhi,gaume2}),
one concludes that in the latter case the ultraviolet finiteness
of a  quantum theory\footnote{According to the analysis of
Ref.~\cite{zanon1}, for the case of two--, and four--dimensional
target the Ricci--flatness condition one reveals at the one--loop
level persists to higher orders in perturbation theory.}
is guaranteed by the presence of a higher symmetry at the classical level.

The organisation of the work is as follows. In the next section we briefly
review the nonlinear sigma models with an $N$--extended global
supersymmetry and discuss the origin of the twisted supersymmetry.
In Sect. 3 the $N=2$ non--linear sigma model and the $N=4$
twisted generalisation are compared in the complex coordinates common to
their string partners. In the former case we reproduce the well known
condition that a manifold must be K\"ahler, while in the latter case it
is shown that in addition it has to admit a covariantly constant holomorphic
two--form in order to support the $N=4$ twisted supersymmetry.
We turn to string theory in Sect. 4 and 5 and consider
gauging of an $N=2$ global supersymmetry and an $N=4$ twisted global
supersymmetry. We explicitly check that in both the cases the gauging does
not impose any new restrictions on the background geometry as compared
to those implied already by the global supersymmetry and hence
is valid to describe a consistent coupling of the $N=2,4$ strings
to the external curved backgrounds.
Calculation of the one-loop $\b$--function is outlined
in Sect. 6. We summarise our results and discuss some further problems in
the concluding Sect. 7. Two Appendices contain our spinor notations in $d=2$
and some elementary notions from complex geometry on K\"ahler manifolds,
these being relevant for the gauging procedure implemented in
Sect. 4,5. Throughout the paper we work in components.
The superfield technique is avoided in order to keep the connection between
geometry and strings at a more transparent level.

\vskip 0.5cm

\noindent
{\bf 2. Nonlinear sigma models and N--extended twisted supersymmetry}\\[-4pt]

\noindent
\vskip 0.4cm

To elucidate the structure of an N--extended twisted supersymmetry
in the sigma model context, it is
worth recalling the original argument by Alvarez--Gaum\'e and
Freedman~\cite{gaume1} valid for ordinary supersymmetry transformations.
Given the free field action on a flat background\footnote{In what follows,
we denote flat $d=2$ vector indices by the letters from the beginning of
the Latin alphabet. Those from the end are reserved for the target space
indices. On a flat background ${e_a}^\a =\d_a^\a$ and
$\g^a {e_a}^\a \equiv \g^\a$.}
\bea\label{toy}
S={\textstyle{\frac 12}}
\int \!\! d^2 x \{ \partial^\a \phi^n \partial_\a \phi_n
-i {\bp}^n \g^\a \partial_\a \p_n \},
\eea
where the target space index $n$ runs from $0$ to $N-1$ and $\g$ are the
Dirac matrices in two dimensions
(for our $d=2$ spinor conventions see Appendix A),
one can easily guess an ansatz for the global supersymmetry
transformation
\bea\label{anz}
\d \phi^n ={f^n}_m (\bar\e \p^m), \quad \d \p^n=\l i{f^n}_m
\partial_\a \phi^m (\g^\a \e).
\eea
The bosonic fields $\phi^n$ and the fermionic partners $\psi^n$
are taken to be real, $\l$ is an arbitrary real constant and
${f^n}_m$ is an arbitrary real nondegenerate (constant) matrix,
both to be determined below.

The invariance of the action under the ansatz leads
${f^n}_m$ to obey (the target space flat metric is denoted by
$\eta_{nm}$)
\be\label{fcond1}
\eta_{pn} {f^n}_m +\l {f^k}_p \eta_{km}=0,
\ee
where one has to take into account the identity
$\g_a \g_b =-\eta_{ab}-\e_{ab}
\g_3$ and integrate by parts. Making use of the $d=2$ Fiertz
identity (see also Appendix A)
\bea\label{fier}
&& (\bar\e_1 \partial_\a \p^n) \g^\a \e_2 =
-(\bar\e_1 \g^\a \e_2) \partial_\a \p^n
-{\textstyle{\frac 12}}(\bar\e_1 \e_2) \g^\a \partial_\a \p^n
-{\textstyle{\frac 12}}(\bar\e_1 \g^\a \partial_\a \p^n)\e_2
\nonumber\\[2pt]
&& \qquad \qquad
+{\textstyle{\frac 12}}(\bar\e_1 \g_3 \e_2)
\g_3 \g^\a \partial_\a \p^n
+{\textstyle{\frac 12}}(\bar\e_1 \g_3 \g^\a \partial_\a \p^n)\g_3\e_2,
\eea
the algebra of the transformations can be readily evaluated (on--shell
relations)
\be\label{salgebra}
[\d_1,\d_2] \phi^n =
i \l {f^n}_m {f^m}_k (\bar\e_2 \g^\a \e_1 -\bar\e_1 \g^\a \e_2)
\partial_\a \phi^k,
\ee
and the same for the field $\p^n$. Thus, for the transformation~(\ref{anz})
to be the standard supersymmetry transformation one has to set
\be\label{fcond2}
\l {f^n}_m {f^m}_k =-{\d^n}_k.
\ee

If one wishes to realize an extended supersymmetry then each $f$ must
satisfy the conditions (\ref{fcond1}), (\ref{fcond2}) with the
corresponding $\l$ involved and besides the vanishing of cross
brackets for two different transformations implies
\bea\label{fcond3}
\l_1 {{(f_2 f_1)}^n}_k +\l_2 {{(f_1 f_2)}^n}_k =0,
\qquad \l_2  {{(f_2 f_1)}^n}_k +\l_1 {{(f_1 f_2)}^n}_k=0.
\eea

Observing further that one of the transformations
can always be generated by the unit matrix
\be\label{fcond4}
{f^n_0}_m={\d^n}_m \longrightarrow \l_0=-1,
\ee
one ends up with the algebraic conditions
\be\label{fcond5}
f^2_i=-1, \qquad \l_i=1, \quad i\ne 0.
\ee
while Eq.~(\ref{fcond3}) reduces to the Clifford algebra (for $i\ne j$).
Notice that on a flat background the number of supersymmetries
does not seem to be bounded from above. It suffices to raise
the value of $N$ and find an appropriate representation of
the Clifford algebra.

The same arguments apply also to a more complicated case of
a curved manifold or a nonlinear sigma model with
an N--extended global supersymmetry. A minimal coupling of
the toy system~(\ref{toy}) to a curved background metric
$g_{nm}(\phi)$ turns out to be not enough to respect the
supersymmetry and some new terms are to be
added~\cite{freed1,gaume1}
\bea
&&
S={\textstyle{\frac 12}}
\int \!\! d^2 x \{ \partial^\a \phi^n \partial_\a \phi^m g_{nm} (\phi)
-i {\bp}^n \g^\a \partial_\a \p^m g_{nm} (\phi)
-g_{nm}(\phi){\G^m}_{pk} (\phi) \partial_\a \phi^p~ i {\bp}^n \g^\a \p^k
\nonumber\\[2pt]
&& \qquad \qquad
-{\textstyle{\frac 16}} R_{nmpk} (\phi) ({\bp}^n \p^p) ({\bp}^m \p^k) \},
\eea
where ${\G^m}_{pk}$ is the Levi--Civita connection
and $R_{nmpk}$ is the Riemann tensor.
Besides, the transformations themselves are to be
slightly modified and ${f^n}_m$ acquires the $\phi$--dependence
\bea\label{anz1}
\d \phi^n ={f^n}_m (\phi) (\bar\e \p^m), \quad \d \p^n=\l i{f^n}_m (\phi)
\partial_\a \phi^m (\g^\a \e) - {\G^n}_{mk}(\phi){f^m}_p (\phi)
(\bar\e \p^p) \p^k.
\eea
Varying the action and verifying the algebra of the transformations one
again comes to the algebraic conditions~(\ref{fcond1}),(\ref{fcond2}),
(\ref{fcond3}),(\ref{fcond4}),(\ref{fcond5}) where the flat metric is to be
exchanged with the curved one. Besides, there appears a new restriction
that ${f^n}_m$ must be covariantly constant
\be\label{fcond6}
\nabla_k {f^n}_m=0.
\ee
Notice that the latter is trivially satisfied for ${f^n_0}_m={\delta^n}_m$
and hence at least one supersymmetry can always be realized without any
restrictions on the background geometry~\cite{freed1}.
In checking the symmetry, the integrability condition (Ricci identity)
\be
R_{nmlp}{f^l}_k-\l R_{nmlk}{f^l}_p=0,
\ee
and the standard properties of the Riemann tensor
\be
{R^n}_{mpk} + \mbox{cycle}(mpk) =0, \quad
\nabla_l {R^n}_{mpk}+ \mbox{cycle}(lpk) =0
\ee
are of heavy use, while the analysis of the (on--shell)
algebra appeals to the fermionic equation of motion
\be
i\g^\a {\mathcal{D}}_\a \p^k +{\textstyle{\frac 13}}
{R^k}_{mpl} \p^p (\bp^m \p^l)=0,
\ee
with ${\mathcal{D}}_\a \p^k =\partial_\a \p^k +{\G^k}_{nm} \partial_\a
\phi^n \p^m$. It is worth noting, that the easiest way to deal with
the last term in the action is to reduce the $d=2$ spinors involved
in it to their irreducible components, like we do in Sect. 4 below
(see also Appendix A).
This allows one to exploit the symmetry properties of the Riemann tensor
more efficiently and liberates one from the necessity to use
Fiertz identities.

In contrast to the flat case, the fact that a target manifold has
to admit a covariantly constant tensor or, in other words,
a tensor commuting with all elements from the holonomy group of a
manifold, restricts severely the number of possible supersymmetries.
The $N\leq 4$ bound has been revealed~\cite{gaume1} for an
irreducible manifold, this based on the Schur's lemma applied to real
representations of the holonomy group. Since a product of two
complex structures necessarily yields a third one (a quaternionic
structure), the K\"ahler
geometry corresponding to $N=2$ and the hyper K\"ahler geometry
associated with $N=4$ seem to be the only options available
\footnote{In Ref.~\cite{BN} geometric models of $N=4$ supersymmetric
mechanics have been proposed,
which can be viewed as one-dimensional 
counterparts of the two-dimensional $N=2$ supersymmetric
sigma-models of \cite{gaume1}.}.

An additional interesting possibility has been brought to the focus
quite recently~\cite{hit,gibb,hull1,hull2}. The observation
made~\cite{hull1} is that for one of the four
transformations\footnote{In what follows we assume that the first
transformation is generated by the unit matrix and the second one
by the ordinary complex structure.},
say $f_2$, one can give up the conventional sign in the commutation
relation~(\ref{salgebra}), thus leading to a twisted supersymmetry
algebra and a "real" structure (or an almost product structure~\cite{yano})
\be\label{twist1}
{f^n_2}_m {f^m_2}_k ={\d^n}_k, \quad \nabla_k {f^n_2}_m =0.
\ee
A product of the latter with the complex structure $f_1$ yields then
another real structure while a product of the two real structures
gives back the complex structure, altogether inducing a
pseudo quaternionic structure~\cite{gibb,hit}. The full algebra is
then an $N=4$
twisted supersymmetry algebra and the corresponding geometry
is known as a pseudo hyper K\"ahler geometry. A detailed discussion
of the latter point in the sigma model context can be found in
Refs.~\cite{hull1,hull2}. Notice that the possibility to twist the
$N=4$ supersymmetry algebra is consistent with Jacobi identities
since for the $d=2$ (twisted) super Poincar\'e algebra
\bea
\{Q_A, Q_B \}= 2 \s {(\g_0 \g^\a)}_{AB} P_\a, \quad
[M, Q_A]={\textstyle{\frac 12}} \g_{3 AB} Q_B, \quad
[M, P_\a]=\e_{\a\b} P^\b,
\eea
those hold both for $\s=1$ and $\s=-1$.

\vskip 0.7cm

\noindent
{\bf 3. K\"ahler and pseudo hyper K\"ahler geometries}\\[-4pt]

\noindent
\vskip 0.4cm

As was mentioned in the Introduction, the prime concern of this work is
to compare the geometry of the $N=2$ string and its $N=4$ topological
extension. Since gauging global (super) symmetries of a sigma model
results in a string theory coupled to background,
it is worth considering the sigma model conformed to the complex
frame intrinsic to the $N=2$ string. The Lagrangian to start
with is that of the $N=2$ string in the superconformal gauge
\be
S=-{\textstyle{\frac {1}{2\pi}}}
\int \!\! d \tau d \s
\{ \partial^\a z^a \partial_\a z^{\bar a} \eta_{a \bar a}
-i{\bar \psi}^{\bar a} \g^\a \partial_\a \psi^a \eta_{a \bar a}
+ i \partial_\a {\bar\psi}^{\bar a} \g^\a \psi^a \eta_{a \bar a} \}.
\ee
Here $z^a$, with $a=0,1$ in the critical dimension,
is a $d=2$ complex scalar and $\psi_A^a$ is
a $d=2$ complex (Dirac) spinor. Since the target space is essentially
complex we shall distinguish between the index carried by a field and
its complex conjugate ${(z^a)}^{*} =z^{\bar a}$, ${(\psi^a)}^{*} =
\psi^{\bar a}$.

It is straightforward to check that the action exhibits the invariance
with respect to an $N=2$ global supersymmetry (the parameter is a
complex spinor)
\be
\d z^a =\bar\e \p^a, \qquad \d \psi^a =-{\textstyle{\frac {i}{2}}}
\partial_\a z^a (\g^\a \e),
\ee
this forming the conventional algebra
\be\label{standcomm}
[\d_1,\d_2]={\textstyle{\frac {1}{2}}}
(i\bar\e_1 \g^\a \e_2 -i\bar\e_2 \g^\a \e_1)
\partial_\a.
\ee
Besides, with a closer look one reveals the invariance under an
$N=2$ twisted supersymmetry
\be
\d z^a =\e_{\bar a \bar b} \eta^{\bar a a} ({\bar\p}^{\bar b} \e), \qquad
\d \p^a ={\textstyle{\frac {1}{2}}} \e_{\bar a \bar b}
\eta^{\bar a a} i \partial_\a z^{\bar b} (\g^\a \e),
\ee
with $\e_{ab}$ the Levi-Civita totally antisymmetric
tensor ($\e_{01}=-1$ and ${(\e_{ab})}^{*}=\e_{\bar a \bar b}$).
The corresponding algebra differs from the standard one~(\ref{standcomm})
by the sign on the right hand side
\be\label{twist2}
[\d^{twist}_1,\d^{twist}_2]=-{\textstyle{\frac {1}{2}}}
(i\bar\e_1 \g^\a \e_2 -i\bar\e_2 \g^\a \e_1)
\partial_\a.
\ee
In verification of Eq.~(\ref{twist2}) the trivial identity
\be
\eta^{\bar a a} \eta^{\bar b c} \e_{\bar a \bar b}=\e^{ac},
\ee
with ${\eta}^{\bar a a}=\mbox{diag}(-,+)$, is helpful.
The cross commutator $[\d,\d^{twist}]$ proves to vanish on--shell.

Turning to a curved background, it seems natural to assume
that the complex structure intrinsic to the flat model
persists in a curved space. A target manifold is thus taken to be
Hermitian with a Hermitian metric $g_{n \bar m} (z,\bar z)$
(we denote the inverse by $g^{\bar m n}$ and
conjugate as ${(g_{n\bar m})}^{*}=g_{m \bar n}$) while
the action functional of the nonlinear sigma model
in this framework acquires the form
\bea\label{sigma2}
&& S=-{\textstyle{\frac {1}{2\pi}}}
\int \!\! d \tau d \s
\{ \partial^\a z^n \partial_\a z^{\bar m} g_{n \bar m}
-i{\bar \psi}^{\bar m} \g^\a \partial_\a \psi^n g_{n \bar m}
-i{\bar \psi}^{\bar m} \g^\a \psi^k~ {\G^n}_{pk} \partial_\a z^p
g_{n \bar m}
\nonumber\\[2pt]
&&
\qquad
+ i \partial_\a {\bar\psi}^{\bar m} \g^\a \psi^n g_{n \bar m}
+i{\bar \psi}^{\bar k} \g^\a \psi^n~ {\G^{\bar m}}_{\bar p \bar k}
\partial_\a z^{\bar p} g_{n \bar m}
+2 R_{\bar m n \bar p k} ({\bar\p}^{\bar m} \p^n)
({\bar\p}^{\bar p} \p^k)
\}.
\eea
Variation of all but last terms in the action under
an $N=2$ global supersymmetry transformation
\be
\d z^n =(\bar\e \p^n), \qquad \d \psi^n =-{\textstyle{\frac {i}{2}}}
\partial_\a z^n (\g^\a \e) -{\G^n}_{mp} (\bar\e \p^m) \p^p,
\ee
yields
\bea\label{var2}
&& (\bar \e \p^k) \partial^\a z^n \partial_\a z^{\bar m}
(\partial_k g_{n \bar m} -\partial_n g_{k \bar m})
+(\bar \e \g_3 \p^n) \partial_\b z^{\bar m}
\e^{\b\a} \partial_\a z^{\bar k} \partial_{\bar k}
g_{n \bar m} \nonumber\\[2pt]
&& -i({\bar \p}^{\bar m} \g^\a \p^p) ({\bar\p}^{\bar k} \e) \partial_\a
z^n R_{\bar m p \bar k n}
-i({\bar \p}^{\bar m} \g^\a \p^p) ({\bar\p}^{\bar n} \e) \partial_\a
z^k R_{p \bar m k \bar n} +\mbox{c.c.}
\eea
Thus for the invariance to hold in a curved action one has to
demand
\be\label{kaehler}
\partial_k g_{n \bar m} -\partial_n g_{k \bar m}=0, \qquad
\partial_{\bar k} g_{n \bar m} -\partial_{\bar m} g_{n \bar k}=0,
\ee
which means that the target Hermitian manifold must be
torsion free or, equivalently,  K\"ahler. On a K\"ahler manifold
the Riemann tensor acquires extra symmetries (for completeness
we list them in Appendix B) which then can be used to show that the
variation of the last term in~(\ref{sigma2}) exactly cancells the remnant
in~(\ref{var2}). Making use of the fermionic equation of motion
one can verify also that the algebra~(\ref{standcomm}) persist
on a K\"ahler space.

Let us now proceed to the $N=2$ twisted transformation.
Because the metric carries the indices of different type,
the naive guess like $\e_{ab} \rightarrow \e_{ab}/\sqrt{ -\mbox{det g}}$,
one could try to implement in passing to a curved space, does not yield
a tensor field. Hence one is forced to introduce an
arbitrary two--form $B_{nm}$, $\e_{nm}$ being the flat limit, and
consider the ansatz
\be
\d z^n = B_{\bar k \bar p} g^{\bar k n}({\bar\p}^{\bar p} \e), \qquad
\d \p^n ={\textstyle{\frac {1}{2}}} B_{\bar k \bar p} g^{\bar k n}
i \partial_\a z^{\bar p} (\g^\a \e)
-{\G^n}_{pm} g^{\bar k p} B_{\bar k \bar p}
({\bar\p}^{\bar p} \e) \p^m.
\ee
With respect to this generalisation the nonlinear sigma
model action holds invariant provided
\bea\label{hypkael1}
&& \nabla_k B_{nm}=0, \qquad \partial_{\bar k} B_{nm}=0,
\nonumber\\[2pt]
&& \nabla_{\bar k} B_{\bar n \bar m}=0, \qquad \partial_k
B_{\bar n \bar m}=0.
\eea
In checking the invariance, the integrability conditions
\be\label{hypkael2}
{R^k}_{n \bar m s} B_{kp} -{R^k}_{n \bar m p} B_{ks} =0,
\qquad {R^{\bar k}}_{\bar n m \bar s} B_{\bar k \bar p}
-{R^{\bar k}}_{\bar n m \bar p} B_{\bar k \bar s} =0,
\ee
prove to be helpful. Evaluating further the algebra,
one encounters one more condition
\be\label{hypkael3}
B_{\bar n \bar m} B_{sp} g^{\bar m s}=g_{p \bar n},
\ee
this however does not seem to be an extra restriction, since
for an irreducible manifold the right hand side must be proportional
to the metric with a constant real coefficient (recall that
both $g$ and $B$ are covariantly constant), the latter can be
obsorbed into a redefinition of $B$~\cite{yano}. Finally,
taking into account the conditions~(\ref{hypkael1}),(\ref{hypkael2})
and the Fiertz identity~(\ref{fier}) one can verify that the
cross commutator $[\d,{\d}^{twist}]$ vanishes on--shell which leads to
the full $N=4$ twisted algebra.

Thus, for the nonlinear sigma model on a K\"ahler space to
admit an extra $N=2$ twisted global supersymmetry, the manifold
must admit a covariantly constant holomorphic two--form.
This means that the holonomy group of a manifold,
which is generally a subgroup of $U(1,1)$, reduces to a subgroup
of $SU(1,1)$ (see for example~\cite{lichn}), the latter point
implies a Ricci--flat space. Actually, contracting the first of the
integrability conditions~(\ref{hypkael2}) with the tensor
$g^{\bar l p} B_{\bar l \bar r} g^{\bar r s}$ one immediately
arrives at
\be
{R^k}_{k\bar n m}=0,
\ee
which is the familiar Ricci--flatness condition.

As is well known, for the $N=2$ model the latter condition appears at
the quantum level as a requirement of the one--loop ultraviolet
finiteness~\cite{mukhi} of the theory. Curiously enough, as we
have seen above, the $N=4$ topological prescription implies it already
in the classical area where it proves to be encoded into a higher symmetry
of the formalism.

\vskip 0.7cm

\noindent
{\bf 4. Gauging $N=2$ global supersymmetry}\\[-4pt]

\noindent
\vskip 0.4cm

We now turn to string theory and consider gauging of the extended
global supersymmetry discussed above.
Given a global symmetry transformation, a conventional way to
convert it to a local one consists in applying the Noether procedure.
In general, extra fields are needed and for the case of
local supersymmetry those usually fill up some or
another $d=2$, $N$--extended conformal supergravity multiplet.
Since gauging the $N=2$ global supersymmetry in the sigma model~(\ref{sigma2})
should result in the $N=2$ string consistently coupled to the curved
background, the structure of the Noether couplings is prompted by
the $N=2$ string itself. For a chiral half, the analysis has been done in
Ref.~\cite{bergnish} while for the ordinary (untwisted)
$N=4$ string the question has been addressed in Ref.~\cite{pern}
(for some related work see~\cite{b3}--\cite{bi}).

In order to avoid cumbersome calculations caused by the
$d=2$ Fiertz rearrangement rules
one has to use in checking the local symmetries, like in Ref.~\cite{bg1}
we choose to get rid of $d=2$ $\gamma$--matrices and work directly
in terms of irreducible components on the world--sheet (for example
$\psi^a_A =\left(\begin{array}{cccc}
\psi^a_{(+)}\\
\psi^a_{(-)}\\
\end{array}\right)$; for our $d=2$ spinor conventions see Appendix A).
The action functional of the $N=2$ gauged nonlinear sigma model is then a
sum of Eq.~(\ref{sigma2}), which we rewrite here in terms of the irreducible
components
\bea\label{gauged1}
&& S_{N=2}=-{\textstyle{\frac {1}{2\pi}}}
\int \!\! d \tau d \s \sqrt{-g}
\{ g^{\a\b} \partial_\a z^n \partial_\b z^{\bar m} g_{n \bar m}
+i\sqrt{2} (\psi^n_{(+)}\partial_\a \psi^{\bar m}_{(+)}+
\psi^{\bar m}_{(+)}\partial_\a \psi^n_{(+)}){e_{-}}^\a
g_{n \bar m}
\nonumber\\[2pt]
&& \qquad \quad +i\sqrt{2} (\psi^n_{(-)}\partial_\a \psi^{\bar m}_{(-)}
+\psi^{\bar m}_{(-)}\partial_\a \psi^n_{(-)}){e_{+}}^\a g_{n \bar m}
+i\sqrt{2}\psi^{\bar m}_{(+)} \psi^s_{(+)} {\G^n}_{ps}
\partial_\a z^p {e_{-}}^\a g_{n \bar m} \nonumber\\[2pt]
&& \qquad
-i\sqrt{2}\psi^{\bar s}_{(+)} \psi^n_{(+)}
{\G^{\bar m}}_{\bar p \bar s}
\partial_\a z^{\bar p} {e_{-}}^\a g_{n \bar m}
+i\sqrt{2}\psi^{\bar m}_{(-)} \psi^s_{(-)} {\G^n}_{ps}
\partial_\a z^p {e_{+}}^\a g_{n \bar m} \nonumber\\[2pt]
&& \qquad
-i\sqrt{2}\psi^{\bar s}_{(-)} \psi^n_{(-)}
{\G^{\bar m}}_{\bar p \bar s}
\partial_\a z^{\bar p} {e_{+}}^\a g_{n \bar m}
+4 R_{\bar n p \bar m k}\psi^{\bar n}_{(+)}
\psi^p_{(-)}
\psi^{\bar m}_{(-)} \psi^k_{(+)}\},
\eea
and a chain of the Noether couplings involving an
$N=2$, $d=2$ world--sheet supergravity multiplet
$({e_a}^\b,\chi_{A \b},A_\b)$ ($a$ stands for a flat index)
\bea\label{gauged2}
&& S_{N=2}^{Noether}=-{\textstyle{\frac {1}{2\pi}}}
\int \!\! d \tau d \s \sqrt{-g} \{
-\sqrt{2} \psi^{\bar m}_{(+)} \psi^n_{(+)} A_\a {e_{-}}^\a
g_{n \bar m} -
\sqrt{2} \psi^{\bar m}_{(-)} \psi^n_{(-)} A_\a {e_{+}}^\a
g_{n \bar m} \nonumber\\[2pt]
&& \qquad  \quad + 2i \partial_\a z^n \psi^{\bar m}_{(-)}
\c_{\b (+)}{e_{-}}^\a {e_{+}}^\b g_{n \bar m}
-2i \partial_\a z^n \psi^{\bar m}_{(+)} \c_{\b (-)}{e_{+}}^\a
{e_{-}}^\b g_{n \bar m} \nonumber\\[2pt]
&& \qquad   \quad
-2i \partial_\a z^{\bar m} {\psi}^n_{(+)} {\bar \c}_{\b (-)}
{e_{+}}^\a {e_{-}}^\b g_{n \bar m}
+2i\partial_\a z^{\bar m}
{\psi}^n_{(-)} {\bar \c}_{\b (+)}{e_{-}}^\a {e_{+}}^\b g_{n \bar m}
\nonumber\\[2pt]
&& \qquad  \quad
-2\psi^{\bar m}_{(+)} \psi^n_{(-)} {\bar \c}_{\a(+)}
\c_{\b (-)} {e_{+}}^\a {e_{-}}^\b g_{n \bar m}
-2\psi^{\bar m}_{(-)} \psi^n_{(+)} {\bar \c}_{\a(-)}
\c_{\b (+)} {e_{+}}^\b {e_{-}}^\a g_{n \bar m}
\nonumber\\[2pt]
&& \qquad  \quad
+ \psi^{\bar m}_{(+)} \psi^n_{(+)} {\bar \c}_{\b(-)} \c_{\a (-)}
({e_{+}}^\a {e_{-}}^\b+{e_{+}}^\b {e_{-}}^\a) g_{n \bar m}
\nonumber\\[2pt]
&& \qquad \quad
+\psi^{\bar m}_{(-)} \psi^n_{(-)} {\bar \c}_{\b(+)} \c_{\a (+)}
({e_{+}}^\a {e_{-}}^\b+{e_{+}}^\b {e_{-}}^\a)g_{n \bar m} \}.
\eea
Because all the terms in a variation of the action which are
proportional either to $\p^{\bar m}_{(+)} \p^n_{(+)}g_{n \bar m}$ or
to $\p^{\bar m}_{(-)} \p^n_{(-)}g_{n \bar m}$ can be compensated
by an appropriate variation of the gauge field $A_\a$, it suffices
to check the invariance modulo those terms. For an $N=2$ local
world--sheet supersymmetry one finds
\bea
&& \d z^n =i {\bar \e}_{(-)} \psi^n_{(+)}, \quad
\d \psi^n_{(-)}=-i{\bar\e}_{(-)} {\G^n}_{pk} \p^p_{(+)}
\p^k_{(-)}, \quad
\d \c_{\a(+)}=0, \quad \d {e_{+}}^\a=0,\nonumber\\
&& \d \psi_{(+)}={\textstyle{\frac {1}{\sqrt{2}}}} \e_{(-)} \partial_\a
z {e_{+}}^\a  -{\textstyle{\frac {i}{\sqrt{2}}}}\e_{(-)}\psi_{(-)}
{\bar\c}_{\g(+)}{e_{+}}^\g
+{\textstyle{\frac {i}{2\sqrt{2}}}}\psi_{(+)}({\bar\e}_{(-)}\c_{\g(-)}-
\e_{(-)} {\bar\c}_{\g(-)}) {e_{+}}^\g,
\nonumber\\[2pt]
&& \d {e_{-}}^\a=
-{\textstyle{\frac {i}{\sqrt{2}}}} {e_{+}}^\a
({\bar\e}_{(-)}\c_{\g(-)}+
\e_{(-)} {\bar\c}_{\g(-)}){e_{-}}^\g,
\nonumber\\[2pt]
&& \d \c_{\a(-)}=\nabla_\a \e_{(-)}
+{\textstyle{\frac {i}{2}}}\e_{(-)}A_\a +
{\textstyle{\frac {i}{2\sqrt{2}}}}
({\bar\e}_{(-)} \c_{\a(-)}+
\e_{(-)} {\bar\c}_{\a(-)})\c_{\g(-)}{e_{+}}^\g\nonumber\\[2pt]
&& \qquad \quad
+{\textstyle{\frac {i}{\sqrt{2}}}}\e_{(-)} \c_{\a(-)} {\bar\c}_{\g(-)}
{e_{+}}^\g,
\eea
and
\bea
&& \d z^n =i {\bar \e}_{(+)} \psi^n_{(-)}, \quad
\d \psi^n_{(+)}=-i{\bar\e}_{(+)} {\G^n}_{pk} \p^p_{(-)} \p^k_{(+)}, \quad
\d \c_{\a(-)}=0, \quad \d {e_{-}}^\a=0, \nonumber\\[2pt]
&& \d \psi_{(-)}={\textstyle{\frac {1}{\sqrt{2}}}} \e_{(+)} \partial_\a
z {e_{-}}^\a  +{\textstyle{\frac {i}{\sqrt{2}}}}\e_{(+)}\psi_{(+)}
{\bar\c}_{\g(-)}{e_{-}}^\g
-{\textstyle{\frac {i}{2\sqrt{2}}}}\psi_{(-)}({\bar\e}_{(+)}\c_{\g(+)}-
\e_{(+)} {\bar\c}_{\g(+)}) {e_{-}}^\g,
\nonumber\\[2pt]
&& \d {e_{+}}^\a=
{\textstyle{\frac {i}{\sqrt{2}}}} {e_{-}}^\a
({\bar\e}_{(+)}\c_{\g(+)}+
\e_{(+)} {\bar\c}_{\g(+)}){e_{+}}^\g,
\nonumber\\[2pt]
&& \d \c_{\a(+)}=-\nabla_\a \e_{(+)}
-{\textstyle{\frac {i}{2}}}\e_{(+)}A_\a -{\textstyle{\frac {i}{2\sqrt{2}}}}
({\bar\e}_{(+)} \c_{\a(+)}+
\e_{(+)} {\bar\c}_{\a(+)})\c_{\g(+)}{e_{-}}^\g
\nonumber\\[2pt]
&& \qquad \quad
-{\textstyle{\frac {i}{\sqrt{2}}}}\e_{(+)} \c_{\a(+)} {\bar\c}_{\g(+)}
{e_{-}}^\g.
\eea
In addition, apart from the usual reparametrization invariance,
local Lorentz transformations and Weyl symmetry,
the model exhibits invariance under two extra bosonic transformations
\be\label{a}
\d A_\a =\partial_\a a, \quad \d \psi_{(\pm)}=-{\textstyle{\frac {i}{2}}}
a \psi_{(\pm)}, \quad \d \c_{(\pm)} =-{\textstyle{\frac {i}{2}}}
a \c_{(\pm)};
\ee
\bea\label{b}
&& \d A_\a =e^{-1} \e_{\a\b} g^{\b\g} \partial_\g b,
\quad \d \psi_{(+)}=-{\textstyle{\frac {i}{2}}}
b \psi_{(+)}, \quad \d \psi_{(-)}={\textstyle{\frac {i}{2}}}
b \psi_{(-)}, \nonumber\\[2pt]
&& \d \c_{(+)}={\textstyle{\frac {i}{2}}}
b \c_{(+)}, \quad \d \c_{(-)}=-{\textstyle{\frac {i}{2}}}
b \c_{(-)},
\eea
where $e^{-1}={(det({e_{n}}^\a))}^{-1}=\sqrt{-g}$,
and the super Weyl transformation
\bea\label{w}
&& \d A_\a={\textstyle{\frac {1}{\sqrt{2}}}} g_{\a\b} {e_{+}}^\b {e_{-}}^\g
({\bar \n}_{(+)}\c_{\g(-)}+{\bar\c}_{\g(-)}\n_{(+)})+
{\textstyle{\frac {1}{\sqrt{2}}}} g_{\a\b} {e_{-}}^\b {e_{+}}^\g
({\bar \n}_{(-)}\c_{\g(+)}+{\bar\c}_{\g(+)}\n_{(-)}),
\nonumber\\[2pt]
&&
\d \c_{\a(+)}=g_{\a\b} {e_{+}}^\b \n_{(-)}, \quad
\d \c_{\a(-)}=g_{\a\b} {e_{-}}^\b \n_{(+)},
\eea
these just preserving their flat form. In checking the local symmetries
one has to use essentially the fact that the target manifold is K\"ahler
and the metric is covariantly constant.
Besides, special care is to be taken of the terms requiring
integration by parts. When integrating by parts,
the derivative $\partial_\a$ will hit the background metric
\be
\partial_\a g_{n\bar m}=\partial_\a z^k \partial_k g_{n\bar m}
+\partial_\a z^{\bar k} \partial_{\bar k} g_{n\bar m}=
\partial_\a z^k {\G^l}_{kn} g_{l \bar m} +
\partial_\a z^{\bar k} {\G^{\bar l}}_{\bar k \bar m} g_{n \bar l},
\ee
thus inducing extra terms as compared to the flat case.

Notice that omitting the world--sheet supergravity fields in the
transformation laws above and taking the parameter to be
a constant, one is left precisely with the
sigma model global supersymmetry transformations, thus supporting the
consistency of the gauging done.

\vskip 0.7cm

\noindent
{\bf 5. Gauging $N=4$ twisted global supersymmetry}\\[-4pt]

\noindent
\vskip 0.4cm
An action functional for the  $N=4$ twisted string in a flat space
has been constructed in Refs.~\cite{bg1,bg2}. Just like we proceeded
in the former case, in order to gauge two remaining twisted
global supersymmetries in the $N=2$ gauged nonlinear sigma model
it suffices to mimic the structure of the terms entering
the $N=4$ twisted string.
To this end, on the world--sheet there must be introduced
two new real vectors ${\cal C}_\a$ and ${\cal D}_\a$, and an extra
complex fermion $\m_{A\a}$~\cite{bg1}, these complementing
an $N=2$, $d=2$ supergravity
multiplet to an $N=4$, $d=2$ one and playing the role of the
gauge fields for the extra local symmetry transformations.
An amendment composed of the new fields reads
\bea\label{s1}
&& S_{N=4}^{Noether}=-{\textstyle{\frac {1}{2\pi}}}
\int \!\! d \tau d \s \sqrt{-g}
\{ \sqrt{2} (\p^n_{(-)} \p^m_{(-)} {e_{+}}^\a+
\p^n_{(+)} \p^m_{(+)} {e_{-}}^\a)({\cal C}_\a +i{\cal D}_\a)B_{nm}
\nonumber\\[2pt]
&& \qquad
+2i\partial_\a z^n \p^m_{(-)} \m_{\b (+)} {e_{-}}^\a {e_{+}}^\b B_{nm}
-2i\partial_\a z^n \p^m_{(+)} \m_{\b (-)} {e_{+}}^\a {e_{-}}^\b B_{nm}
\nonumber\\[2pt]
&& \qquad
-{\textstyle{\frac {1}{2}}}
\p^n_{(+)}\p^m_{(+)} \m_{\a(-)} (\c+\bar\c)_{\b(-)}
g^{\a\b} B_{nm}
-{\textstyle{\frac {1}{2}}}
\p^n_{(-)}\p^m_{(-)} \m_{\a(+)} (\c+\bar\c)_{\b(+)}
g^{\a\b} B_{nm}
\nonumber\\[2pt]
&& \qquad
+2\p^n_{(-)} \p^m_{(+)}({\bar\c}_{\a(+)} \m_{\b(-)}+{\bar\c}_{\b(-)}
\m_{\a(+)}) {e_{+}}^\a {e_{-}}^\b B_{nm}
\nonumber\\[2pt]
&& \qquad
+2\p^n_{(+)} \p^{\bar m}_{(-)} {\bar\m}_{\a (+)}
\m_{\b(-)}{e_{+}}^\a {e_{-}}^\b g_{n\bar m} +\mbox{c.c.}~ \}.
\eea
Before we display a local form of the twisted supersymmetry
transformations, it is worth verifying that the adding of the
further Noether couplings we gathered above in
$S_{N=4}^{Noether}$ to the previous action $S_{N=2}+S_{N=2}^{Noether}$
does not destroy the local symmetries
intrinsic to the $N=2$ gauged nonlinear sigma model. For this to be
the case, the local $\e_{(-)}$--transformation is to include
an extra piece in the variation of the  field $\p^n_{(+)}$
and besides the fermionic gauge field $\m_{\b(-)}$ has to be
involved too
\bea
&&
\d \p^n_{(+)} =-{\textstyle{\frac {i}{\sqrt{2}}}} B_{\bar k \bar s}
g^{\bar k n} \e_{(-)} \p^{\bar s}_{(-)} {\bar \m}_{\g (+)}{e_{+}}^\g,
\quad {e_{+}}^\b \d \m_{\b(+)} =0, \nonumber\\[2pt]
&&
{e_{-}}^\b \d \m_{\b(-)} = i\e_{(-)} {({\cal C}+i{\cal D})}_\b
{e_{-}}^\b +{\textstyle{\frac {i}{\sqrt{2}}}} \m_{\b(-)}
({\bar\e}_{(-)}\c_{\g(-)}+\e_{(-)} {\bar\c}_{\g(-)})
{e_{-}}^\g {e_{+}}^\b
\nonumber\\[2pt]
&&
-{\textstyle{\frac {i}{2\sqrt{2}}}} \e_{(-)} \m_{\g(-)}
{(\c+\bar\c)}_{\b(-)} g^{\g\b}
-{\textstyle{\frac {i}{2\sqrt{2}}}}\m_{\b (-)}
 ({\bar\e}_{(-)} \c_{\g(-)}
-\e_{(-)} {\bar\c}_{\g(-)}){e_{+}}^\g {e_{-}}^\b.
\eea
Similarly, for the $\e_{(+)}$--transformation one finds
\bea
&&
\d \p^n_{(-)} ={\textstyle{\frac {i}{\sqrt{2}}}} B_{\bar k \bar s}
g^{\bar k n} \e_{(+)} \p^{\bar s}_{(+)} {\bar \m}_{\g (-)}{e_{-}}^\g,
\quad
{e_{-}}^\b \d \m_{\b(-)} =0,
\nonumber\\[2pt]
&&
{e_{+}}^\b \d \m_{\b(+)} = -i\e_{(+)} {({\cal C}+i{\cal D})}_\b
{e_{+}}^\b -{\textstyle{\frac {i}{\sqrt{2}}}} \m_{\b(+)}
({\bar\e}_{(+)}\c_{\g(+)}+\e_{(+)} {\bar\c}_{\g(+)})
{e_{+}}^\g {e_{-}}^\b
\nonumber\\[2pt]
&&
+{\textstyle{\frac {i}{2\sqrt{2}}}} \e_{(+)} \m_{\g(+)}
{(\c+\bar\c)}_{\b(+)} g^{\g\b}
+{\textstyle{\frac {i}{2\sqrt{2}}}}\m_{\b (+)}
({\bar\e}_{(+)} \c_{\g(+)}
-\e_{(+)} {\bar\c}_{\g(+)}){e_{-}}^\g {e_{+}}^\b.
\eea
Because all the terms in a variation of the full action
which are proportional either to
$\p^n_{(+)} \p^m_{(+)} B_{nm}$ or to $\p^n_{(-)} \p^m_{(-)} B_{nm}$
can be compensated by an appropriate variation of the gauge field
${\cal C}_\a +i{\cal D}_\a$ and the same is obviously true for
complex conjugates, it suffices to check the invariance modulo
those terms. Besides, the transformation law of the field $A_\a$ which by the
same reason we omitted in the previous section will
be modified by new contributions involving $\m_{(\pm)}$.

Turning to the transformations~(\ref{a}),(\ref{b}),(\ref{w}),
one discovers that the following contributions from the extra gauge
fields
\bea
&&
\d_a \m_{\b(+)}={\textstyle{\frac {i}{2}}} a \m_{\b(+)}, \quad
\d_a \m_{\b(-)}={\textstyle{\frac {i}{2}}} a \m_{\b(-)},
\nonumber\\[2pt]
&&
\d_a {({\cal C} +i{\cal D})}_\a =i a {({\cal C} +i{\cal D})}_\a
-a {\textstyle{\frac {i}{2\sqrt{2}}}} \m_{\g(-)}
\c_{\b(-)} g^{\g\b} {e_\a}^{-}
\nonumber\\[2pt]
&& \qquad
-a {\textstyle{\frac {i}{2\sqrt{2}}}} \m_{\g(+)}
\c_{\b(+)} g^{\g\b}{e_\a}^{+};
\eea
\bea
&&
\d_b \m_{\b(+)}=-{\textstyle{\frac {i}{2}}} b \m_{\b(+)}, \quad
\d_b \m_{\b(-)}={\textstyle{\frac {i}{2}}} b \m_{\b(-)},
\nonumber\\[2pt]
&&
\d_b {({\cal C} +i{\cal D})}_\a =i b
\{ {({\cal C} +i{\cal D})}_\g  {e_{-}}^\g
-{\textstyle{\frac {1}{2\sqrt{2}}}} \m_{\g(-)}
\c_{\b(-)} g^{\g\b} \} {e_\a}^{-}
\nonumber\\[2pt]
&& \qquad
-i b \{ {({\cal C} +i{\cal D})}_\g {e_{+}}^\g
-{\textstyle{\frac {1}{2\sqrt{2}}}} \m_{\g(+)}
\c_{\b(+)} g^{\g\b} \} {e_\a}^{+};
\eea
and
\bea
\d_\n {({\cal C} +i{\cal D})}_\a ={\textstyle{\frac {1}{2\sqrt{2}}}}
\m_{\g(-)} {(\n+\bar\n)}_{(+)} {e_{-}}^\g {e_\a}^{-}
+{\textstyle{\frac {1}{2\sqrt{2}}}}
\m_{\g(+)} {(\n+\bar\n)}_{(-)} {e_{+}}^\g {e_\a}^{+},
\eea
render the action invariant when
combined with~(\ref{a}),(\ref{b}) and (\ref{w}).

Having completed the consistency check, we now proceed to discuss
the twisted local supersymmetry in the full action
\be
S_{N=4}=S_{N=2}+S_{N=2}^{Noether}+S_{N=4}^{Noether}.
\ee
After an extremely tedious calculation with the extensive use
of Eqs.~(\ref{kaehler}),(\ref{hypkael1})--(\ref{hypkael3}),
one can verify that the action holds invariant under the $N=2$ twisted
local supersymmetry with a fermionic parameter $\k_{(+)}$
\bea
&& \d z^n = B_{\bar k \bar p} g^{\bar k n}
i \psi^{\bar p}_{(-)} \k_{(+)},
\quad
\d \c_{\a(-)}=0, \quad \d \m_{\a(-)}=0, \quad \d {e_{-}}^\a =0,
\nonumber\\[2pt]
&&
\d \psi^n_{(+)}=-{\G^n}_{pk} g^{\bar k p} B_{\bar k \bar p}
i\p^{\bar p}_{(-)} \k_{(+)} \p^k_{(+)},
\quad
\d {e_{+}}^\a=
{\textstyle{\frac {i}{\sqrt{2}}}}
(\k_{(+)}\m_{\g(+)} +{\bar\k}_{(+)} {\bar\m}_{\g(+)})
{e_{+}}^\g {e_{-}}^\a,
\nonumber\\[2pt]
&&
{e_{+}}^\b \d \c_{\b (+)} =-i{\bar\k}_{(+)} {({\cal C} -i{\cal D})}_\a
{e_{+}}^\a-{\textstyle{\frac {i}{\sqrt{2}}}}(\k_{(+)}\m_{\g(+)}+
{\bar\k}_{(+)}{\bar\m}_{\g(+)}) \c_{\b(+)}{e_{-}}^\b {e_{+}}^\g
\nonumber\\[2pt]
&& \qquad\quad
-{\textstyle{\frac {i}{2\sqrt{2}}}}
{\bar\k}_{(+)}{\bar\m}_{\g(+)}{(\c+\bar\c)}_{\b(+)}g^{\b\g},
\nonumber\\[2pt]
&& \d \psi^n_{(-)}={\textstyle{\frac {1}{\sqrt{2}}}}
B_{\bar k \bar p} g^{\bar k n} \partial_\a z^{\bar p} \k_{(+)}
{e_{-}}^\a -{\textstyle{\frac {i}{\sqrt{2}}}} \p^n_{(+)} \k_{(+)}
\m_{\g(-)}{e_{-}}^\g
-{\textstyle{\frac {i}{\sqrt{2}}}}
B_{\bar k \bar p} g^{\bar k n} \p^{\bar p}_{(+)} \k_{(+)} \c_{\g(-)}
{e_{-}}^\g
\nonumber\\[2pt]
&& \qquad \quad
-{\G^n}_{pk} g^{\bar k p} B_{\bar k \bar p} i\p^{\bar p}_{(-)} \k_{(+)}
\p^k_{(-)},
\nonumber\\[2pt]
&& {e_{+}}^\b \d \m_{\b (+)} = \nabla_\b {\bar\k}_{(+)} {e_{+}}^\b
-{\textstyle{\frac {i}{2}}} {\bar\k}_{(+)} A_\b {e_{+}}^\b
-{\textstyle{\frac {i}{2\sqrt{2}}}} {\bar\k}_{(+)} {\bar\c}_{\b(+)}
\c_{\g(+)} g^{\g\b} -{\textstyle{\frac {i}{\sqrt{2}}}}
(\k_{(+)}\m_{\g(+)}
\nonumber\\[2pt]
&& \qquad \quad
+{\bar\k}_{(+)} {\bar\m}_{\g(+)}) \m_{\b(+)}
{e_{+}}^\g {e_{-}}^\b,
\eea
and $\k_{(-)}$
\bea
&& \d z^n = B_{\bar k \bar p} g^{\bar k n}
i \psi^{\bar p}_{(+)} \k_{(-)},
\quad
\d \c_{\a(+)}=0, \quad \d \m_{\a(+)}=0, \quad \d {e_{+}}^\a =0,
\nonumber\\[2pt]
&&
\d \psi^n_{(-)}=-{\G^n}_{pk} g^{\bar k p} B_{\bar k \bar p}
i\p^{\bar p}_{(+)} \k_{(-)} \p^k_{(-)},
\quad
\d {e_{-}}^\a=-
{\textstyle{\frac {i}{\sqrt{2}}}}
(\k_{(-)}\m_{\g(-)} +{\bar\k}_{(-)} {\bar\m}_{\g(-)})
{e_{-}}^\g {e_{+}}^\a,
\nonumber\\[2pt]
&&
{e_{-}}^\b \d \c_{\b (-)} =i{\bar\k}_{(-)} {({\cal C} -i{\cal D})}_\a
{e_{-}}^\a+{\textstyle{\frac {i}{\sqrt{2}}}}(\k_{(-)}\m_{\g(-)}+
{\bar\k}_{(-)}{\bar\m}_{\g(-)}) \c_{\b(-)}{e_{+}}^\b {e_{-}}^\g
\nonumber\\[2pt]
&& \qquad\quad
+{\textstyle{\frac {i}{2\sqrt{2}}}}
{\bar\k}_{(-)}{\bar\m}_{\g(-)}{(\c+\bar\c)}_{\b(-)}g^{\b\g},
\nonumber\\[2pt]
&& \d \psi^n_{(+)}={\textstyle{\frac {1}{\sqrt{2}}}}
B_{\bar k \bar p} g^{\bar k n} \partial_\a z^{\bar p} \k_{(-)}
{e_{+}}^\a +{\textstyle{\frac {i}{\sqrt{2}}}} \p^n_{(-)} \k_{(-)}
\m_{\g(+)}{e_{+}}^\g
+{\textstyle{\frac {i}{\sqrt{2}}}}
B_{\bar k \bar p} g^{\bar k n} \p^{\bar p}_{(-)} \k_{(-)} \c_{\g(+)}
{e_{+}}^\g
\nonumber\\[2pt]
&& \qquad \quad
-{\G^n}_{pk} g^{\bar k p} B_{\bar k \bar p} i\p^{\bar p}_{(+)} \k_{(-)}
\p^k_{(+)},
\nonumber\\[2pt]
&& {e_{-}}^\b \d \m_{\b (-)} =- \nabla_\b {\bar\k}_{(-)} {e_{-}}^\b
+{\textstyle{\frac {i}{2}}} {\bar\k}_{(-)} A_\b {e_{-}}^\b
+{\textstyle{\frac {i}{2\sqrt{2}}}} {\bar\k}_{(-)} {\bar\c}_{\b(-)}
\c_{\g(-)} g^{\g\b} +{\textstyle{\frac {i}{\sqrt{2}}}}
(\k_{(-)}\m_{\g(-)}
\nonumber\\[2pt]
&& \qquad \quad
+{\bar\k}_{(-)} {\bar\m}_{\g(-)}) \m_{\b(-)}
{e_{-}}^\g {e_{+}}^\b.
\eea
Here again we omitted rather lengthy expressions for the variations
of the world--sheet vector fields $A_\a$ and ${\cal C}_\a +i{\cal D}_\a$,
these being responsible for removing the terms proportional to
$\p^n_{(\pm)} \p^{\bar m}_{(\pm)}
g_{n\bar m}$ and $\p^n_{(\pm)} \p^m_{(\pm)} B_{nm}$.
A relevant technical point to mention is that in verification of the
$\kappa_{(\pm)}$-invariance it proves to be helpful to cancel
the terms in the following sequence: first the terms involving $\nabla \p$
and its complex conjugate, then those containing $\partial z$
and $\partial \bar z$,
then the terms quadratic in $\partial z$, $\partial \bar z$, and then the rest.
This is because the integration by parts in the $\nabla \p$--terms
will contribute to $\partial z$, ${\partial z}^2$ and so on.

Apart from the transformations listed above, the previous work on the
structure of the $N=4$ topological string in a flat space~\cite{bg1,bg2}
indicates the presence of further local symmetries. They prove
to involve two complex bosonic parameters and two complex fermionic ones
which together with $\kappa_{(\pm)}$
match perfectly the number of the extra gauge fields on the world--sheet.
It is straightforward to check that the transformations
persist in the curved space just maintaining their flat form.
Omitting the variations of $A_\a$ and ${\cal C}_\a +i{\cal D}_\a$,
the bosonic pair can be represented as
\bea
&&
\d_c \p^n_{(+)}=c B_{\bar m \bar p} g^{\bar m n} \p^{\bar p}_{(+)},
\quad \d_c \p^n_{(-)}=c B_{\bar m \bar p} g^{\bar m n} \p^{\bar p}_{(-)},
\quad \d_c \c_{\a(+)} =-c\m_{\a(+)},
\nonumber\\[2pt]
&&
\d_c \c_{\a(-)} =-c\m_{\a(-)},
\quad \d_c \m_{\a(+)} =-\bar c \c_{\a(+)},
\quad \d_c \m_{\a(-)} =-\bar c \c_{\a(-)},
\eea
and
\bea
&&
\d_f \p^n_{(+)}=f B_{\bar m \bar p} g^{\bar m n} \p^{\bar p}_{(+)},
\quad \d_f \p^n_{(-)}=-f B_{\bar m \bar p} g^{\bar m n} \p^{\bar p}_{(-)},
\quad \d_f \c_{\a(+)} =f\m_{\a(+)},
\nonumber\\[2pt]
&&
\d_f \c_{\a(-)} =-f\m_{\a(-)},
\quad \d_f \m_{\a(+)} =\bar f \c_{\a(+)},
\quad \d_f \m_{\a(-)} =-\bar f \c_{\a(-)},
\eea
while for the fermionic transformation one finds
\be
\d \m_{\a (-)}=\l_{(+)} g_{\a\b} {e_{-}}^\b, \quad
\d \m_{\a (+)}=\l_{(-)} g_{\a\b} {e_{+}}^\b.
\ee
That the $N=4$ twisted global supersymmetry can be
gauged without imposing further restrictions on the background
geometry implies the consistency of the coupling and provides
one with the action functional of the $N=4$ twisted string
coupled to a Ricci--flat K\"ahler background.

\vskip 0.7cm

\noindent
{\bf 6. One--loop $\b$--function}\\[-4pt]
\noindent
\vskip 0.4cm

As we have seen above, the action functional to describe the 
$N=4$ twisted nonlinear sigma model is identical to that of the 
$N=2$ theory. Then the structure of one--loop ultraviolet divergences 
in the $N=4$ model is immediately elucidated due
to the analysis available for the $N=2$ case  
(see e.g.~\cite{mukhi}). For the completeness of the 
presentation we mention here a few relevant points.

When analysing ultraviolet behaviour of a theory,
it is customary to use the background field method (for a review
see Ref.~\cite{zachos}). To maintain manifest covariance in 
perturbation theory, it is convenient to switch to normal 
coordinates (for the details and conventions see~\cite{zachos})
\bea
&&
S [\rho(x,s=1), \psi(x)]=\sum_{n=1}^{\infty}
\left. \frac{1}{n!}\frac{d^n}{ds^n}S[\rho(x,s), \p(x)]\right|_{s=0}=
\nonumber\\[2pt] 
&&
=\sum \left. \frac{1}{n!}D^n(s)S[\rho(x,s), \p(x)]\right|_{s=0}
=S_0 + S_1 + S_2 + \ldots~,
\eea
where $S_0$ is given by~(\ref{sigma2}), the argument
being the background field. Splitting the metric in the common way
$g_{n\bar m}=V_n^a V_{\bar m}^{\bar b} \eta_{a\bar b}$
and redefining the quantum field ${\textstyle{\frac {d~\rho^n}{d~s}}}{|}_{s=0}
\rightarrow V^a_n{\textstyle{\frac {d~\rho^n}{d~s}}}{|}_{s=0}$ one ends
up with the usual framework of quantum field theory, the
propagators and vertices being easy to define.  

Turning to one--loop divergences, one reveals that
a potentially divergent contribution from the fermionic fields 
involves the integral (in momentum space)
\begin{eqnarray}
\int d^dp\frac{2p^{\alpha}p^{\beta}-\delta^{\alpha\beta}p^2}
{((p+k)^2-m^2)(p^2-m^2)}+(finite~part~as~d\rightarrow 2),
\end{eqnarray}
which proves to vanish as the divergent contribution from 
the first and the second terms 
exactly cancel each other. Notice that this is in a perfect
agreement with the absence of an ultraviolet
divergence in the self-energy of a minimally coupled vector potential
in two dimensions~\cite{honer2}. The same result can be confirmed
working in superfields~\cite{mukhi,dk}.

Then a detailed analysis shows~\cite{mukhi}
that the structure of one--loop divergences is specified
completely by the $S_2$ vertex in a sector of the bosonic fields  
\begin{eqnarray}
-\frac{1}{2\pi}\int d^2\sigma R_{n\bar m k\bar l}
\partial^{\alpha} z^k \partial_{\alpha}z^{\bar l}
<~\xi^n~\xi^{\bar m}~>,
\end{eqnarray}
the divergent part thus involving
\begin{eqnarray}
-\frac{1}{4\pi\epsilon}{R^k}_{k\bar n m}
\partial^{\alpha}z^m\partial_{\alpha}z^{\bar n}.
\end{eqnarray}
Beautifully enough,
as the classical $N=4$ twisted model is formulated on 
a Ricci--flat K\"ahler manifold, 
one immediately concludes that the corresponding quantum
theory is automatically free of ultraviolet divergences at 
the one--loop level. 

\vspace{0.7cm}

\noindent
{\bf 7. Conclusion}\\[-4pt]
\noindent
\vskip 0.4cm
To summarise, in the present paper we have compared the nonlinear
sigma model possessing an $N=2$ global supersymmetry with its
$N=4$ twisted generalisation. The extra twisted transformations
were constructed with the use of a background two--form field.
We argued that in order to provide a symmetry of the
$N=2$ action, the two--form must be covariantly constant and holomorphic.
This is known to reduce the holonomy group
to a subgroup of $SU(1,1)$ and implies a Ricci--flat K\"ahler background.
Gauging of both
the $N=2$ and $N=4$ global (twisted) supersymmetries has been
performed appealing to the $N=2$, $d=2$ and $N=4$, $d=2$ supergravity
multiplets on the world--sheet, respectively. Recalling further the fact
that the string partners of the sigma models are physically equivalent
in a flat space, and that the $N=4$ extension has the advantage of being
manifestly Lorentz invariant,  it seems tempting to speculate
that the latter point is responsible for the improved ultraviolet
behaviour of the $N=4$ twisted theory.

Turning to possible further developments, it would be interesting
to derive the K\"ahler condition on the metric and those on the
two--form field directly from the $N=2,4$ superconformal algebra
in a curved space.
The correct form of the superconformal currents is prompted by the
gauged versions of the sigma models we constructed above.
Since in the Hamiltonian framework the currents appear as secondary
constraints, it is far from obvious that the information following from the
closure of the algebra on a background will be as restrictive as that
implied by the local Lagrangian symmetries. Although we suspect they
should match. Another interesting point is the superfield version
of the analysis undertaken in this work.

\vskip 0.6cm

\noindent
{\bf Acknowledgements}\\[-4pt]

\noindent

The work of two of us (S.B. and A.G.) has been supported by
INTAS grant No 00 OPEN 254 and Iniziativa Specifica
MI12 of the Commissione IV of INFN.

\vspace{0.6cm}

\noindent
{\bf Appendix A}\\[-4pt]

\noindent

In this Appendix we gather our $d=2$ spinor notations and discuss
some technical points relevant for the verification of local symmetry
transformations of the $N=2$ string and its $N=4$ twisted extension
coupled to external curved background.

In order to describe fermions on the world--sheet of a string,
it is customary to use purely imaginary
$\g$--matrices
$$
\begin{array}{lll}
\g_0=\left(\begin{array}{cccc}
0 & -i\\
i & 0\\
\end{array}\right), \qquad
\g_1=\left(\begin{array}{cccc}
0 & i\\
i & 0\\
\end{array}\right), \qquad
\g_3=\g_0 \g_1 =\left(\begin{array}{cccc}
1 & 0\\
0 & -1\\
\end{array}\right).
\end{array}
\eqno{(A.1)}
$$
These obey the algebraic properties
$$
\begin{array}{lll}
\{ \gamma_a, \gamma_b \} =-2\eta_{ab}, \quad
\gamma_a \gamma_b =-\eta_{ab} -\epsilon_{ab} \g_3,
\quad \e^{ab} \g_b =\g_3 \g^a,
\end{array}
\eqno{(A.2)}
$$
with $\eta_{ab} ={\it diag} (-,+)$ and $\epsilon_{ab}$ the 2d Levi-Civita
totally antisymmetric tensor, $\epsilon_{01}=-1$. The second and the third
identities are
specific to the two dimensional space and simplify dealing with the
$\g$--matrices considerably. Notice further that
the charge conjugation matrix $C$, $\g_a^{T} =-C\g_a C^{-1}$, just
coincides with $\g_0$ and furthermore
${(\gamma_0 \gamma_a)}^{+}=\g_0 \g_a$,
${(\gamma_0 \gamma_a)}^{T}=\g_0 \g_a$, where ${\dots}^{+}$ stands for the
Hermitian conjugation and ${\dots}^{T}$ for the transposition.

Any $2\times 2$ complex matrix $M_{AB}$ can be decomposed with
respect to the basis $\{ 1_2, \g_a, \g_3 \}$
$$
\begin{array}{lll}
M_{AB} =a \d_{AB} +a_b \g^b_{AB} +b\g_{3 AB},
\end{array}
\eqno{(A.3)}
$$
the coefficients having the form
$$
\begin{array}{lll}
&& a={\textstyle{\frac {1}{2}}} Tr( M), \quad a_b =
-{\textstyle{\frac {1}{2}}} Tr( M\g_b), \quad b=
{\textstyle{\frac {1}{2}}} Tr( M \g_3).
\end{array}
\eqno{(A.4)}
$$
Taking $M_{AB}=\p_A \vf_B$, with arbitrary spinors $\p_A$,
$\vf_B$, and differentiating with respect to
the latter one gets the basic Fiertz identity
$$
\begin{array}{lll}
\d_{AK} \d_{BN} ={\textstyle{\frac {1}{2}}} \d_{AB} \d_{KN}
-{\textstyle{\frac {1}{2}}} \g^b_{AB} \g_{b NK}
+{\textstyle{\frac {1}{2}}} \g_{3 AB} \g_{3 NK}.
\end{array}
\eqno{(A.5)}
$$
This allows one to rearrange the order of spinors in various
expressions involving world--sheet fermions.
The key identity which proves to be of extensive use in checking
the local supersymmetry of the $N=2$ string action is
$$
\begin{array}{lll}
(\bar\p \vf) (\bar\c \g_3 \l)-(\bar\p \g_3 \vf) (\bar\c \l)=
(\bar\c \g^b \vf) (\bar\p \g_b \g_3 \l),
\end{array}
\eqno{(A.6)}
$$
while the one helpful in verification of global  ${U(1,1)}_{outer}$
invariance reads
$$
\begin{array}{lll}
(\bar\p \vf) (\bar\c \g_3 \l)+(\bar\p \g_3 \vf) (\bar\c \l)=
-(\bar\p \l) (\bar\c \g_3 \vf)-(\bar\p \g_3 \l) (\bar\c \vf).
\end{array}
\eqno{(A.7)}
$$

Since in $d=2$ irreducible representations of the Lorentz group
are one--dimensional, it is sometimes convenient to use the
light-cone notation for vectors and spinors
$$
\begin{array}{lll}
&& A_{\pm}={\textstyle{\frac {1}{\sqrt 2}}}(A_0 \pm A_1),
\quad A^n B_n = -A_{+} B_{-} -A_{-} B_{+}, \quad
\Psi_A =\left(\begin{array}{cccc}
\psi_{(+)}\\
\psi_{(-)}\\
\end{array}\right),\nonumber\\[2pt]
&&
\g_{-}=-i\sqrt{2}\left(\begin{array}{cccc}
0 & 1\\
0 & 0\\
\end{array}\right), \qquad
\g_{+}=i\sqrt{2}\left(\begin{array}{cccc}
0 & 0\\
1 & 0\\
\end{array}\right), \nonumber\\[2pt]
&& A^{+}=-A_{-}, \quad A^{-}=-A_{+}, \quad
\g_3 \g_{+}=-\g_{+}, \quad  \g_3 \g_{-}=\g_{-},
\end{array}
\eqno{(A.8)}
$$
in which the Lorentz transformations simplify to
$$
\begin{array}{lll}
\d A_{\pm} =\pm \L A_{\pm}, \quad \d \psi_{(\pm)}=\pm
\textstyle{\frac 12} \L\psi_{(\pm)}=\textstyle{\frac 12} \L \g_3 \psi.
\end{array}
\eqno{(A.9)}
$$
Obviously, the invariance is kept by contracting a ``${+}$'' with
a ``${-}$'' (one could fairly well contract a ``${(+)}$'' with a ``${(-)}$''
or a ``${+}$'' with two ``${(-)}$'').
When a local version of the Lorentz transformations is considered,
a spin connection $\o_\a$ is to be introduced, this allowing one to
construct covariant derivatives
$$
\begin{array}{lll}
&&
\nabla_\a \p_{(+)}=\partial_\a \p_{(+)} -\o_\a \p_{(+)}, \quad
\nabla_\a \p_{(-)}=\partial_\a \p_{(-)} +\o_\a \p_{(-)},
\nonumber\\[2pt]
&&
\nabla_\a A_{+}=\partial_\a A_{+} -2\o_\a A_{+}, \quad
\nabla_\a A_{-}=\partial_\a A_{-} +2\o_\a A_{-},
\end{array}
\eqno{(A.10)}
$$
with $\d \o_\a=\textstyle{\frac 12} \partial_\a \L$.
The advantage of the light--cone notation is that it allows one to get
rid of $\gamma$--matrices and work explicitly in terms of the irreducible
components of tensors under consideration. Notice that taking into
account the properties of the charge conjugation matrix it is easy to
check that the object $\p^a C \g^k \vf^b \e_{ab}$ transforms as a complex
vector under the $SO(1,1)$ Lorentz group, while $\p^a C \vf^b \e_{ab}$
is a complex scalar.

Introducing the zweibein ${e_b}^\a$ on the world--sheet
$g^{\a\b}={e_b}^\a \eta^{bc} {e_c}^\b $ and its inverse ${e_\a}^b$,
$ g_{\a\b}= {e_\a}^b \eta_{bc} {e_\b}^c$, where $\a$ stands for a curved
index, one can finally verify the relations
$$
\begin{array}{lll}
{e_n}^\a {e_k}^\g -{e_n}^\g {e_k}^\a =e \e^{\a\g} \e_{kn},
\quad
g^{\a\g} {e_k}^\b -g^{\a\b} {e_k}^\g =e \e^{\g\b} \e_{kn} e^{n\a}.
\end{array}
\eqno{(A.11)}
$$
These prove to be helpful in checking the local supersymmetry of the
$N=2$ string. Other useful identities which we extensively use in the text
are
$$
\begin{array}{lll}
\e^{cp} \e^{sk} =-\eta^{cs} \eta^{pk} + \eta^{ck} \eta^{ps},
\quad \e^{mn}  {e_m}^\a {e_n}^\b =e \e^{\a\b}, \quad
\e_{\a\b}  {e_n}^\a {e_m}^\b =e \e_{nm},
\end{array}
\eqno{(A.12)}
$$
where $e=det({e_n}^\a)$ and $\e_{\a\b}$ is a totally antisymmetric matrix
with $\e_{01}=-1$.

Turning to the light--cone framework, some identities relevant to this work
are
$$
\begin{array}{lll}
&& g^{\a\b}=-{e_{+}}^\a {e_{-}}^\b -{e_{-}}^\a {e_{+}}^\b, \quad
e \epsilon^{\a\b}= -{e_{+}}^\a {e_{-}}^\b +{e_{-}}^\a {e_{+}}^\b,
\nonumber\\[2pt]
&&
g_{\a\b}=-{e_{\a}}^{+} {e_\b}^{-} -{e_\a}^{-} {e_\b}^{+}, \quad
e^{-1} \epsilon_{\a\b}= {e_\a}^{+} {e_\b}^{-} -{e_\a}^{-} {e_\b}^{+},
\nonumber\\[2pt]
&& {e_{-}}^\a {e_\a}^{-}=1, \quad {e_{+}}^\a {e_\a}^{+}=1,
\quad {e_{-}}^\a {e_\a}^{+}=0, \quad {e_{+}}^\a {e_\a}^{-}=0,
\nonumber\\[2pt]
&& \eta_{++}=\eta_{--}=0, \quad \eta_{+-}=\eta_{-+}=-1,
\quad \e_{+-}=1, \quad \e_{-+}=-1.
\end{array}
\eqno{(A.13)}
$$

\vspace{0.7cm}

\noindent
{\bf Appendix B}\\[-4pt]

\noindent
In order to make the presentation self--contained, in this Appendix
we list symmetry properties of the Riemann tensor on a K\"ahler manifold.
These prove to be of heavy use both in verification of
an $N$--extended global supersymmetry for the sigma model under
consideration and in establishing the local version of the
latter.

Given a complex manifold with a Hermitian metric $g_{n\bar m}$
(we denote the inverse by $g^{\bar m n}$), one introduces covariant
derivatives
$$
\begin{array}{lll}
&& \nabla_n v_m =\partial_n v_m -{\G^k}_{nm} v_k, \qquad
\nabla_{\bar n} v_m =\partial_{\bar n} v_m,
\nonumber\\[2pt]
&& \nabla_{\bar n} v_{\bar m} =\partial_{\bar n} v_{\bar m}
-{\G^{\bar k}}_{\bar n \bar m} v_{\bar k}, \qquad
\nabla_n v_{\bar m} =\partial_n v_{\bar m}.
\end{array}
\eqno{(B.1)}
$$
Assuming the covariant constancy of the metric
$$
\begin{array}{lll}
&& \nabla_n g_{m\bar k} =\partial_n g_{m \bar k}
-{\G^p}_{nm} g_{p \bar k}=0, \qquad
\nabla_{\bar n} g_{m\bar k} =\partial_{\bar n} g_{m \bar k}
-{\G^{\bar p}}_{\bar n \bar k} g_{m \bar p}=0,
\end{array}
\eqno{(B.2)}
$$
one readily finds the explicit form of the Levi--Civita connection
$$
\begin{array}{lll}
&& {\G^k}_{nm} = g^{\bar p k} \partial_n g_{m \bar p}, \quad
{\G^{\bar p}}_{\bar n \bar k} = g^{\bar p m} \partial_{\bar n} g_{m \bar k}.
\end{array}
\eqno{(B.3)}
$$
The simplest way to define the curvature and the torsion tensors
is to consider a commutator of two covariant derivatives.
For example
$$
\begin{array}{lll}
[\nabla_A, \nabla_B] v^p ={R^p}_{kAB} v^k -T^D_{AB} \nabla_D v^p,
\end{array}
\eqno{(B.4)}
$$
where $A$ is a collective notation for $a$ and $\bar a$, and
the sum over D involves both $d$ and $\bar d$.
A simple inspection of the latter relation
with the use of Eq.~(B.3) shows that the only non
vanishing components of the torsion tensor are
$$
\begin{array}{lll}
T^k_{nm}={\G^k}_{nm} -{\G^k}_{mn}=g^{\bar k k}(\partial_n g_{m\bar k}-
\partial_m g_{n \bar k}), \quad
T^{\bar k}_{\bar n \bar m}={\G^{\bar k}}_{\bar n \bar m}
-{\G^{\bar k}}_{\bar m \bar n}=g^{\bar k k}(\partial_{\bar n} g_{k\bar m}-
\partial_{\bar m} g_{k \bar n}),
\end{array}
\eqno{(B.5)}
$$
while those of the curvature tensor are exhausted by
$$
\begin{array}{lll}
{R^k}_{n \bar p m}=-{R^k}_{n m \bar p}=\partial_{\bar p} {\G^k}_{mn},
\quad
{R^{\bar k}}_{\bar n p \bar m}=-{R^{\bar k}}_{\bar n \bar m p}=
\partial_p {\G^{\bar k}}_{\bar m \bar n}.
\end{array}
\eqno{(B.6)}
$$
Introducing the notation
$$
\begin{array}{lll}
R_{\bar k n \bar p m}=g_{a\bar k} {R^a}_{n \bar p m},
\end{array}
\eqno{(B.7)}
$$
and making use of the explicit form of the connection, one finds
$$
\begin{array}{lll}
R_{\bar n m \bar p k}=-R_{m \bar n \bar p k}.
\end{array}
\eqno{(B.8)}
$$
Assuming finally that the manifold at hand is a K\"ahler space
$$
\begin{array}{lll}
T^k_{nm}=0 \rightarrow {\G^k}_{nm}={\G^k}_{mn} \rightarrow
\partial_n g_{m\bar k}-
\partial_m g_{n \bar k}=0,
\end{array}
\eqno{(B.9)}
$$
one immediately reveals an extra symmetry for the Riemann tensor
$$
\begin{array}{lll}
{R^k}_{n m \bar p}={R^k}_{m n \bar p}.
\end{array}
\eqno{(B.10)}
$$
Being combined with those valid for an arbitrary Hermitian manifold
the latter yields a chain of relations
$$
\begin{array}{lll}
R_{\bar n m \bar p k}=R_{\bar p m \bar n k}=R_{\bar n k \bar p m}=
R_{\bar p k \bar n m},
\end{array}
\eqno{(B.11)}
$$
which are known as the cyclic property of the Riemann tensor on a
K\"ahler manifold.

Finally, when dealing with the term involving the
Riemann tensor which enter the sigma model action
the following Bianchi identities
$$
\begin{array}{lll}
\nabla_k R_{\bar n m \bar p l}-\nabla_m R_{\bar n k \bar p l}=0,
\quad
\nabla_{\bar k} R_{\bar n m \bar p l}-\nabla_{\bar n} R_{\bar k m \bar p l}=0,
\end{array}
\eqno{(B.12)}
$$
which are valid for a K\"ahler space, prove to be helpful.

\end{document}